\documentclass{jaa}
\usepackage{natbib}
\usepackage{graphicx}

\usepackage{url}

\usepackage[colorlinks=true,citecolor=blue]{hyperref}
\begin{document}\sloppy

\title{In-situ acceleration of radio-emitting particles in the lobes of radio\\  galaxies: Evolving observational perspective and recent clues}


\author{Gopal-Krishna\textsuperscript{1} and Paul J.\ Wiita\textsuperscript{2}}
\affilOne{\textsuperscript{1}UM-DAE Centre for Excellence in Basic Sciences (CEBS), Vidyanagari, Mumbai-400098, India\\}
\affilTwo{\textsuperscript{2}Department of Physics, The College of New Jersey, 2000 Pennington Rd., Ewing, NJ 08628-0718, USA\\}


\twocolumn[{

\maketitle

\corres{wiitap@tcnj.edu}

\msinfo{12 December 2023}{15 February 2024}

\begin{abstract}
The issue of radiation mechanisms had triggered in 1950--60s the first applications of plasma physics to understand the nature of radio galaxies.  This interplay has steadily  intensified during the past five decades, due to the premise of {\it in-situ} acceleration of relativistic electrons occurring in the lobes of radio galaxies. This article briefly traces the chain of these remarkable developments, largely from an observational perspective. We recount several observational and theoretical milestones established along the way and the lessons drawn from them. We also present a new observational clue about {\it in-situ} acceleration of the relativistic particles radiating in the lobes of radio galaxies, gleaned by us from the very recently published sensitive radio observations of a tailed radio source in the galaxy cluster Abell 1033.

\end{abstract}


}]


\doinum{12.3456/s78910-011-012-3}
\artcitid{\#\#\#\#}
\volnum{000}
\year{0000}
\pgrange{1--}
\setcounter{page}{1}
\lp{1}

\section{Introduction}
\label{sec:int}

The story of (non-solar) radio astronomy began with the description as ``radio stars'' of the first few newly discovered discrete radio sources (soon identified with comparatively nearby  optical galaxies), thanks to the poor angular resolving power afforded by the existing radio telescopes which operated at metre wavelengths (Bolton 1956 and references therein). These findings spawned the birth of relativistic astrophysics, when the radio emission from galaxies  was (correctly) identified as synchrotron radiation (Pikelner 1953; Shklovskii 1953; 1955),  and the equipartition argument was invented to quantify their (stupendous) energy contents (Burbidge 1956). 

The era of radio galaxies began in earnest with the discovery at Jodrell Bank (U.K.), by Jennison \& Das Gupta (1953), of a twin-lobed radio morphology of the bright radio galaxy Cygnus A;  they found the radio emission to come from 
two `lobes'  separated by 1.47 arc-minutes, straddling,   roughly symmetrically,  the optical galaxy but clearly separated from it. A decade later, it was shown at Stanford (U.S.A.) that the two ``well-separated'' lobes of Cygnus A are actually connected by a continuous ``bridge'' of radio emission (Swarup, Thompson \& Bracewell 1963), as recounted in a recent historical note (Gopal-Krishna 2021). In fact, until around the mid-1970s the development towards understanding the radio galaxy phenomenon largely remained hinged to Cygnus A, chiefly due to the coincidental location of this extended and extremely powerful, hence  cosmologically rare, radio galaxy in proximity to our Galaxy.

Early   extensive reviews of the phenomenon of radio galaxies and their morphological subclasses were presented by Moffet (1966), Willis (1978),
and Miley (1980). The first theoretical proposals to explain the morphology and/or confinement of the lobes of radio galaxies include those of De Young \& Axford (1967), Christiansen (1969), and Mills \& Sturrock (1970). The edge-brightened shape of the lobes, predicted in some of these works, was confirmed in several independent observations (Mitton \& Ryle 1969 and references therein). Subsequently, such edge-brightened twin-lobed radio sources came to be known as Fanaroff-Riley type II (FR II, Fanaroff \& Riley 1974) and the issue of spectral index gradients across such {\it classical double radio sources} is one key theme traced in this paper\footnote{Prior to 1970, the most comprehensive information about structures of radio galaxies came using the Cambridge one-mile telescope (Ryle 1962), which also yielded spectral gradients across some sources (Macdonald et al.\ 1968). However, the scope of this information was quite limited, due to the small frequency range covered (408--1407 MHz) and the  large beam size ($\sim 80$ arcsec).}

The next major milestone on the morphological track was again set using Cygnus A. It was based on the 5-km aperture-synthesis radio telescope operating from Cambridge (U.K.), often at 11 and 6 cm wavelengths and providing a resolution of 2 arc-seconds at 6 cm (Ryle 1972). These new radio maps clearly showed one or two bright compact regions of size $\sim 1$ kpc located near the outer edge of each  lobe.  These were termed ``hot spots'', and their radio brightness and locations were argued to support the need for a {\it continuous} energy supply from the central galaxy via two oppositely-directed collimated channels (or ``beams''), which terminate and deposit energy in the (advancing) hot spot of each lobe (Hargrave \& Ryle 1974), as theorised by Rees (1971), Longair, Ryle \& Scheuer (1973), Scheuer (1974), and Blandford \& Rees (1974).

The high-pressure synchrotron plasma, containing ultra-relativistic electrons accelerated in the hot spots was envisioned to continually escape from there and flow back towards the parent galaxy, the ``backflow'' (Scheuer 1974; Blandford \& Rees 1974). A solid confirmation of the postulated narrow ``beams'' of energy supply in Cygnus A came from the detection of a bi-polar pair of radio ``jets'', one in each lobe, using the Very Large Array (VLA) telescope in New Mexico (U.S.A.) (e.g., Perley et al.\ 1984). Such bi-polar twin-jets have since become an established paradigm and these ubiquitous channels are now firmly believed to transport energy 
from the central engine to the outer extremities of the lobes  of FR II sources (e.g., Bridle \& Perley 1984; Begelman, Blandford \& Rees 1984). From geometric arguments based on the observed lobe-length asymmetry, it has been estimated that the hot spots marking the jet's termination, typically advance at a speed of ~ 0.01c, or somewhat higher (Scheuer 1995; Arshakian \& Longair 2000). For a normal-sized radio galaxy, spanning ~ 100 -- 300 kpc, such speeds imply a typical age between $10^7$ and $10^8$ years. This {\it kinematical} estimate of the  typical age of radio galaxies is a critical parameter for quantifying their role in several important cosmic phenomena (e.g., Fabian 2012; Blandford et al.\ 2019 and references therein). An independent check on this age estimation is possible by measuring spectral index variations along the lobes of radio galaxies (see below) which are believed to grow continually as the hot spot moves ahead. A critical assessment of this and several other issues arising from the various asymmetries exhibited by classical double radio sources is presented by Gopal-Krishna \& Wiita (2005).

\section{Searches for spectral index gradients along the radio lobes}

The identification of the terminal hot spot as the primary injection site of synchrotron plasma into the radio lobe behind it, arising from the predicted ``backflow'' of the synchrotron plasma towards the parent galaxy was, in principle, amenable to observational verification. Besides checking  for the very existence of radio tails behind the hot spots, one could also look in well-resolved radio maps for signatures of radiative losses suffered by the radiating particles deposited by the advancing hot spots along the tail/lobe. Expectedly, these losses would produce a progressive steepening of radio spectrum, from the hot spot towards the central parent galaxy. This trend was clearly observed in the uniquely bright and extended double radio source Cygnus A through aperture-synthesis mapping at 2.7 and 5 GHz using the Cambridge 1-mile radio telescope (Mitton \& Ryle 1969), and later in more granular detail by mapping Cygnus A at the same two frequencies with the newly commissioned Cambridge 5-km radio telescope (Hargrave \& Ryle 1974). 
These results  predicted the radio tails behind the hot spots becoming increasingly more prominent towards longer wavelengths.

However, verifying generality of this effect was contingent upon the availability of arc-second resolution with good sensitivity at metre wavelengths. This dual requirement came within reach using lunar occultation observations made with the large steerable cylindrical radio telescope built at Ooty (India), to operate at 327 MHz with an effective collecting area of $\sim 8000$ m$^2$ (Swarup et al. 1971)\footnote{The technique of lunar occultation for determining positions and brightness profiles of celestial radio sources was originally proposed by Getmantsev \& Ginzberg (1950). It came into  the limelight when its application by Hazard, Mackay \& Shimmins (1963) paved the way to the discovery of the first quasar, 3C 273 (Schmidt 1963). The restoration technique for recovering the strip-brightness distribution from a lunar occultation profile, which required the removal of the effects of Fresnel diffraction at the moon's limb, was first developed by Scheuer (1962),
 von Hoerner (1964), and Cohen (1969). 
Early applications of this technique to radio galaxies and quasars are summarised in 
Taylor \& De Jong (1968). With the exception of 
3C 273, the occultations were observed using medium-size telescopes (the 140-ft or 85-ft dishes in U.S.A.) and covering a narrow frequency range (200 -- 500 MHz). Due to these constraining factors, these pioneering occultation observations yielded quite limited information on spectral index gradients across radio galaxies.   Likewise, no significant information on spectral index gradients could be adduced  from  lunar occultation observations of 6 sources at 81.5 MHz (Collins \& Scott 1969).  The same holds for the occultation of 3C 212 reported at the very low frequency of 20--25 MHz (Bovkun 1976).}  
The  special design of the Ooty telescope 
made it the instrument of choice for observing lunar occultations. The Ooty telescope became operational in 1970 and  soon thereafter, with the commissioning of the Cambridge 5-km radio telescope (Ryle 1972), mapping with arcsecond resolution at much higher frequencies of 5 and 15 GHz, also became  feasible. (Note that a high resolution at centimetre wavelengths is also crucial for minimising/mitigating the contamination of the lobe's brightness distribution by the jet, which   often becomes significant at such short wavelengths).

The availability of arc-second resolution at both centimetre and metre wavelengths
opened  up the prospect of making detailed comparison of structures of classical double radio sources across a truly large spectral range (albeit, the sources were chosen by the Moon). The 6 sources for which such a comparison could be made were 3C 172 (Jenkins \& Scheuer 1976) and 3C33, 3C 79,  3C 139.2, 3C 154, \& 3C 215 (Gopal-Krishna \& Swarup 1977). For another bright radio galaxy 3C 192, a broad comparison of structures at 256 MHz (Taylor \& De Jong 1968) and 2.7 GHz (Harris 1973) was made by the latter author who noted a general  similarity  between its appearances at the two frequencies. Out of all these sources, only for 3C 215 (a quasar in a cluster at $z$ = 0.411) were the tails/bridge found to become much more prominent (relative to the hotspots) at metre wavelength; for the remaining 6 sources, any such trend was found to be modest, at best. This rather unexpected finding indicated occurence of {\it in-situ} (i.e., spatially distributed) acceleration of relativistic particles in the tails of at least some classical double radio sources (Gopal-Krishna \& Swarup 1977), as was also inferred for the head-tail  radio galaxy NGC 1265 in the Perseus cluster  (Pacholczyk \& Scott 1976). For another three classical double radio sources, namely 3C 219 (Turland 1975), the giant 3C 326 (Willis \& Strom 1978) and 3C 382 (Burch 1979), aperture-synthesis maps covering a narrow frequency range (2.7--5 GHz), showed a near-absence of spectral index variation along at least one of the two radio lobes, which these authors interpreted as  evidence for {\it in-situ} acceleration of relativistic electrons across the lobes.

\section{Dawn of the age of spectral ageing}

In a reversal of the above tactic, measurements of spectral gradients soon began to be exploited as a tool for estimating the ages of radio galaxies of different radio morphological flavours. However, this approach operated under a slew of simplifying assumptions, including the supposition that no {\it in-situ} particle acceleration occurs within the lobes of double radio sources, which are moreover taken to be uniformly filled with the synchrotron plasma and not subject to significant expansion losses  (e.g., Burch 1979; H{\"o}gbom 1979; Winter et al.\ 1980). Here we recall an influential early paper, by van der Laan \& Perola (1969), in which an upper age limit of $\sim10^7$ years was deduced for bright radio galaxies, from the observed near-absence of spectral steepening (up to 5 GHz) in the extensive and homogeneous set of radio measurements of flux densities of nearly 300 3C radio sources (Kellermann et al.\ 1969). The novelty of this age limit lies in its independence from the actual strength of magnetic field in the radio lobes where relativistic particles lose their energy continuously via synchrotron radiation and also by inverse Compton interactions with the cosmic microwave background photons. Remarkably, their inferred upper limit of  $\sim 10^7$ years for the duration of the radio-emitting phase is several times smaller than the typical (statistical) lifetime which a massive elliptical galaxy is believed to spend in that phase, as estimated by Schmidt (1966) from the statistics of identification of bright radio sources with massive elliptical galaxies. 

This discrepancy led van der Laan \& Perola (1969) to hypothesize that radio-emitting particles in the lobes leak away continually, on a time scale comparable to their radiative lifetime in the lobes and, as a result of such bulk diffusion of radiating particles, the lobe emission over the entire frequency range gets attenuated. One expected that the putative particle diffusion would manifest as broadening of the lobes/bridge with increasing wavelength of observation. However, an early check revealed no such trend (Gopal-Krishna 1977), based on a comparison of the available lunar occultation observations at metre wavelengths of the radio galaxies 3C 132 
and 3C 192, with their published maps made at 2.7/5 GHz with the Cambridge 5-km telescope (Gopal-Krishna 1977 and references therein).
Further, it was pointed out in that paper that the existence of a prominent bridge, having a very steep spectrum, in the quasar 3C 215,  as inferred from its lunar occultation observations at 327 MHz (Gopal-Krishna \& Swarup 1977), was not at all hinted by its integrated spectrum, which  remains  straight between 10 MHz and 15 GHz.
This was an early lesson that mere straightness of radio spectrum of a classical double radio source over a wide frequency range does not exclude  the presence of a prominent radio bridge of very steep spectrum in the source. This point acquires added salience in the context of a later finding that a straight radio spectrum is a common feature of those classical double radio sources whose spectra at decimeter wavelengths are ultra-steep (Mangalam \& Gopal-Krishna 1995; Klamer et al. 2006). 

The method of estimating spectral ages of classical double radio sources has grown in both popularity and sophistication, although it still remains firmly rooted in the basic theory developed by Kardashev (1962), Kellermann (1966), Pacholczyk (1970), Jaffe \& Perola (1973), and Miley (1980). Starting from the pioneering studies by Willis and Strom (1978), Burch (1979), H{\"o}gbom (1979), Winter et al (1980), and van Breugel (1980), there is a long and still growing list of innovative works on this topic. These studies include: Myers \& Spangler (1985), Alexander \& Leahy (1987), 
Wiita \& Gopal-Krishna (1990), Carilli et al.\ (1991), Liu et al.\ (1992), Tribble (1993), Katz-Stone \& Rudnick (1997), Eilek et al.\ (1997), Murgia et al.\ (1999), Jones et al.\ (1999), Blundell \& Rawlings (2000), Machalski et al.\ (2007), O'Dea et al.\ (2009), Hardcastle (2013), Harwood et al. (2017), and Mahatma et al.\ (2020).

Expectedly, such analyses acquire greater precision in the case of giant radio galaxies whose lobes span more than a megaparsec (e.g., Gopal-Krishna et al.\ 1989;  Ishwara-Chandra \& Saikia 1999; Jamrozy et al.\ 2008). This is because radiative losses of the synchrotron plasma filling their lobes are dominated by inverse-Compton interactions with the cosmic microwave background photons whose energy density at a given redshift is both accurately known and distributed uniformly. The first quantitative estimation of the role of such inverse-Compton losses in causing ``reduced radio efficiency (RRE)", and thus curtailing the abundance of giant radio galaxies towards higher redshifts was reported in 
Gopal-Krishna et al.\ (1989).\footnote{For a more general account of this, sometimes 
termed ``youth - redshift degeneracy", see Blundell \& Rawlings (1999).} This was based on an analytical treatment of the jet dynamics which also allowed for relativistic speeds of the hotspots. Such ultrafast-moving hot spots  may well exist in double-double radio sources, where a pair of currently energised inner lobes are roughly collinear with the pair of fading outer lobes (see Konar \& Hardcastle 2013; Safouris et al.\ 2008). Note that our 1989 model was a generalisation of our preceding analytical model of dynamic evolution of classical double radio sources (Gopal-Krishna \& Wiita 1987), wherein  a realistic ambient gas density profile was considered, for the first time, as revealed by X-ray observations of massive earty-type galaxies with the Einstein observatory (Nulsen et al.\ 1984). Analytical modelling of the dynamical evolution of high-power radio galaxies, with additional improvements, has since been performed in many studies (e.g., Gopal-Krishna \& Wiita 1991; Kaiser \& Alexander 1997; Blundell et al.\ 2002; Barai \& Wiita 2006;
Kawakatu et al.\ 2009; Turner \& Shabala 2015).

\section{Claims of {\it in-situ} particle acceleration in radio lobes: Twists and turns}

Claims of {\it in-situ} particle acceleration occurring in the lobes of double radio sources of different morphological types (section 2) have kept re-surfacing, notwithstanding the various pitfalls involved in using the spectral gradients along the lobe axis to deduce the age of the electrons radiating in their different parts. Some of the key assumptions made in spectral ageing analysis are: (i) the particle acceleration primarily occurs within the hot spots; (ii) lobes are uniformly filled with synchrotron plasma; (iii) any expansion losses either have an ignorable impact on the age estimation, or they can be accounted for, to the first order; and (iv) within the lobes, electron populations of differing ages do not mix significantly (for additional caveats, see, e.g.,  Blundell \& Rawlings 2001; Rudnick et al.\ 1994; Eilek \& Arendt 1996; Harwood et al.\ 2017). One factor that plays a critical role in spectral age determination is the adopted value of the spectral index $\alpha_{\rm inj}$ of the synchrotron plasma injected from the hot spot into the tail/lobe region,  and the problems in estimating this empirically using the hot spots (e.g., Alexander 1987; Carilli et al. 1991; Machalski et al.\ 2007). Recent analyses incorporating high-resolution LOFAR images near 100 MHz cast some doubt on the traditional practice of taking the slope of the integrated spectrum at such low frequencies as $\alpha_{\rm inj}$ (e.g., McKean et al. 2016; Harwood et al. 2017; Mahatma et al. 2020). A related point is that $\alpha_{\rm inj}$ may even depend on factors external to the source, such as the ambient density (Athreya \& Kapahi 1998); however, the available empirical evidence does not appear to support this  assertion (see Gopal-Krishna et al.\ 2012; Konar \& Hardcastle 2013). Further, it has been empirically claimed that more powerful jets yield a steeper $\alpha_{\rm inj}$ (Konar \& Hardcastle 2013), a correlation also predicted from modelling of dissipation of the jet upon entering the hot spot region (Gopal-Krishna \& Wiita 1990; Ekejiuba et al.\ 1994). All this underscores the criticality of improving our understanding of the particle injection spectrum for  enhancing the precision of the spectral age estimation.

Instinctively, the need for {\it in-situ} particle acceleration would appear particularly pressing in cases where the trend of progressive spectral steepening, from the hot spot towards the centre, is seen to reverse, even if only over a limited region. A prominent early example of this was observed in the classical double radio source 3C 234 at $z = 0.1848$, which has a radio extent of $\sim 108^{\prime\prime}$ ($\sim 330$ kpc for H$_0 = 70$ km~s$^{-1}$ Mpc$^{-1}$).  From a detailed analysis of high-resolution images at five well-separated radio frequencies, Alexander (1987) determined the age distribution for the radiating plasma along its well-aligned pair of radio lobes.
Unexpectedly, that analysis showed the estimated age to remain constant over the first half of the western lobe which is closer to the host galaxy (i.e., no sign of ageing over a $\sim 100$ kpc region, as opposed to the outer, i.e., younger, half of that lobe). The author interpreted this peculiar spectral pattern as evidence for a distributed stochastic re-acceleration of relativistic electrons occurring, probably via turbulence, in the first (older) half of the western lobe. This perceived need for 
{\it in-situ} re-acceleration, however, was shown to be unnecessary in an analysis by Wiita \& Gopal-Krishna (1990) who invoked, for the first time, a (mild and plausible) large-scale spatial gradient of magnetic field along lobes of radio galaxies. Note that their explanation for the anomalous spectral gradient works primarily because the  magnetic energy density in that region of the lobe seems to approach that of the cosmic microwave background.

 Electron reacceleration by galaxy cluster shocks has also been shown to be necessary in some cases, for instance in the merging galaxy cluster Abell 3411--3412 (van Weeren et al.\ 2017).  They used a combination of X-ray, optical and multifrequency radio data from GMRT and VLA  that provide spectral index maps, to show that a tailed radio galaxy in this cluster was attached to an extensive peculiar region of diffuse radio emission, often called a fossil (e.g., En{\ss}lin \& Gopal-Krishna 2001).  This connection indicates that radio galaxies provide both the energetic electrons and the underlying acceleration mechanism to produce that otherwise enigmatic extended relic emission when that plasma is compressed by the shocks in the intracluster medium that are produced in the cluster merger.  A recently discovered radio relic in the cluster Abell 2065  provides additional good evidence for the seeding of the remnant radio emission by an AGN in the cluster (Lal 2021).

\begin{figure*}
\begin{center}
\includegraphics[width=16 cm]{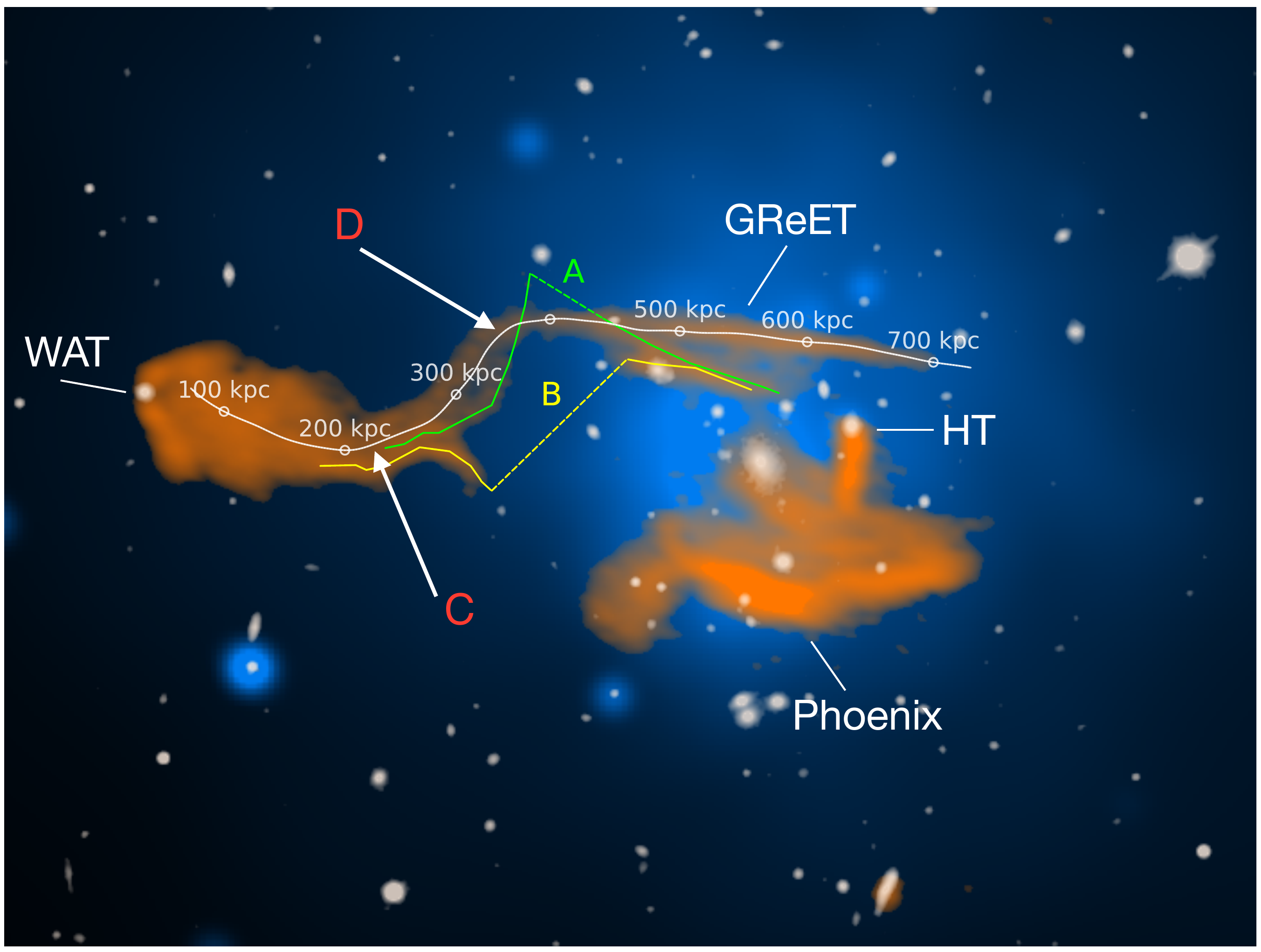}
\caption{A composite image of Abell 1033 from a LOFAR 144 MHz map (orange), the Chandra X-ray image (blue) and the SDSS g-band (white), reproduced with permission from Edler et al.\ (2022; CC-BY 4.0 license).  The white line traces the path used to analyse the spectral properties of the continuous, and gently reenergized, tail (GReET) of the Wide Angle Tail (WAT).   A second head-tail radio galaxy in the field is denoted HT and an irregular steep spectrum source is labelled a Phoenix, because it is probably reenergized by adiabatic compression of an old radio lobe via a shock wave.  Paths A (green) and B (yellow) indicate the location of possible interrupted radio structures.  
}
\label{Fig1}
\end{center}
\end{figure*}

\begin{figure*}
\begin{center}
\includegraphics[width=9cm]{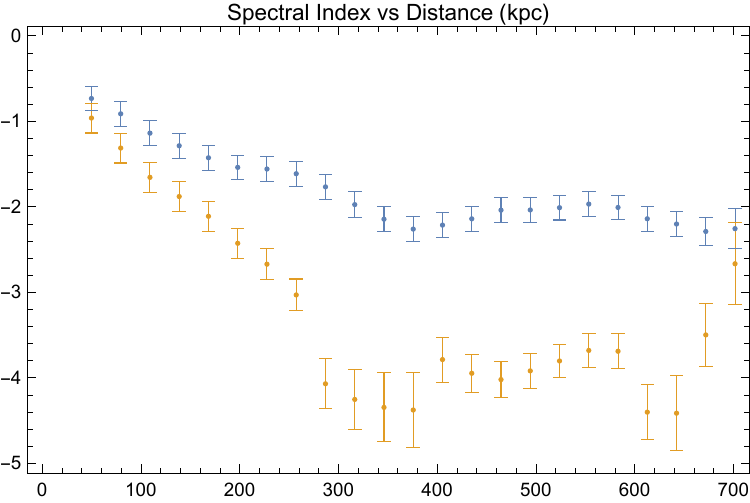}
\caption{Spectral indices (blue) between LOFAR maps at 54 and 144 MHz, and (orange) between 144 MHz (LOFAR) and 323 MHz (GMRT). Data provided by H.\ Edler from Fig.\ 4 of Edler et al.\ (2022; CC-BY 4.0 license).}
\label{Fig2}
\end{center}
\end{figure*}

\subsection{Some recent indicators and clues}

During the past two decades or so, X-shaped radio galaxies (XRGs) have come into the limelight as the testbeds for the astrophysical phenomenon of {\it in-situ} particle acceleration in the lobes of radio galaxies (e.g., Rottmann 2001; Dennet-Thorpe et al. 2002; Lal \& Rao 2005,  2007; Lal et al. 2019; Gopal-Krishna \& Dabhade 2022; Patra et al. 2023). As the name suggests, XRGs exhibit two non-collinear pairs of radio lobes, one of which usually appears edge-brightened and is therefore called the ``primary'' lobes. The other lobe pair is, as a rule, edge-darkened and usually of lower surface brightness, hence called ``secondary''. The secondary lobes (also called ``wings'') are widely thought to be the relics of past nuclear activity in the host galaxy when the twin-jets had been pointing in the direction of the wings, i.e., until the axis of the central engine flipped abruptly to the new direction towards the observed primary lobe pair (e.g., Rottmann 2001; Zier \& Biermann 2001; Merritt \& Ekers 2002). 

In an alternative visualisation, the two wings could merely represent bulk diversion of the synchrotron plasma of the two primary lobes, streaming backwards and impinging 
on the hot gaseous halo of the parent elliptical galaxy (e.g., Leahy \& Williams 1984; 
Capetti et al.\ 2002; Cotton et al.\ 2020; Gopal-Krishna \& Dabhade 2022). Both these paradigms, together with some other possibilities are discussed in a review article  on XRGs by Gopal-Krishna et al.\  (2012). In a radically different approach, the two lobe pairs  in XRGs are created by two (unresolved) active central engines located within the stellar core of the parent galaxy (Lal \& Rao 2005; Lal et al.\ 2019). This scenario, however, fails to account for the absence of a terminal hot spot in all the secondary lobes observed thus far, as well as the recent finding that the radio spectra of the wings are almost never found to be flatter than those of the primary lobes (Patra et al.\ 2023).

These empirical facts notwithstanding, there does exist one robust example, namely the XRG 3C 223.1, where the wings do exhibit a distinctly flatter radio spectrum compared to the primary lobes, even compared to their hot spots. This is seen from a recent high-precision spectral imaging, based on its  LOFAR map at 144 MHz 
and the VLA maps at much higher frequencies going up to 8 GHz (Gopal-Krishna \& Dabhade 2022).  
This rare, counter-intuitive result, based on an exceptionally potent combination of angular resolution  and wavelength coverage (1:58), leaves little doubt, and strongly corroborates the original suggestion by Rottmann\ (2001), which however was based on maps of 3C 223.1 with a modest angular resolution and sensitivity. A similar trend was hinted in subsequent studies of this XRG (Dennett-Thorpe et al.\ 2002; Lal \& Rao 2005).

Recently, Gopal-Krishna \& Dabhade (2022) have underscored the ``double boomerang'' type radio morphology of this XRG 3C 223.1, similar to that found by Cotton et 
al.\ (2020) in their spectacular MeerKAT maps of the giant XRG PKS 2014$-$55. The former authors argue that the origin of this morphology probably lies in the ``bouncing'' of the stream of back-flowing synchrotron plasma of the primary lobes, as it  impinges on an extended (and possibly magnetised) gaseous disk, which is already known to be  associated with the parent elliptical galaxy. Thus, Gopal-Krishna \& Dabhade (2022) propose that the spectral flattening  so distinctly observed towards the wings in 3C 223.1 is probably caused due to particle acceleration (or re-acceleration) triggered during the sharp rebound of the backflowing synchrotron plasma stream. As supporting evidences, they cite the detection of spectral flattening near the bouncing point of (i) the back-flowing synchrotron plasma of the northern primary lobe of the XRG PKS 2014-55 (see Fig.\ 5 in Cotton et al.\ 2020), and (ii) in the northern jet of the Wide-Angle-Tail (WAT) radio galaxy MRC 0600$-$399 (termed ``double-scythe"), as seen in its MeerKAT map at 1.28 GHz (Fig.\ 1 of Chibueze et al.\ 2021). In both these radio galaxies, localised regions of enhanced radio emission, accompanied by spectral flattening, are actually observed near the region where a powerful collimated flow of synchrotron plasma (of a jet or backflow) undergoes a sharp bending or rebound. The  MHD simulations reported in Chibueze et al.\ (2021) have shown that when a jet  carrying synchrotron emitting plasma is deflected upon encountering the tension of external magnetic field lines, an efficient conversion of the magnetic energy into relativistic particles via magnetic reconnection can occur. Then the relativistic particles accelerated {\it in-situ} get transported along the deflected stream of synchrotron plasma. In some cases, this process may be effective enough to force a detectable spectral flattening in the lobes downstream of the deflection point (see also the relativistic MHD simulations by Giri et al.\ 2022).

The couple of examples available thus far, recounted in the foregoing paragraph, serve to boost optimism about pinpointing the physical circumstances and mechanism behind {\it in-situ} acceleration of relativistic particles in the lobes of radio galaxies. Therefore, it is vital to find at least a few more well-founded examples of this curious and long-debated astrophysical phenomenon. From the current observational perspective, we believe that the clue related to bending of a collimated flow of synchrotron plasma holds considerable promise, because of its being amenable to direct observations. Below we draw attention to a radio galaxy in which bending of the lobe/tail is revealing fairly distinct signatures of triggering {\it in-situ} particle acceleration/re-acceleration.

\subsection{The Wide-Angle-Tail radio galaxy in the cluster Abell 1033: A crucial new catch }

The radio galaxy we now discuss is a Wide-Angle-Tail (WAT) associated with a merging galaxy cluster Abell 1033 at $z$ =  0.126. Properties of this system, together with its imaging observations at 54 MHz (LOFAR), 144 MHz (LOFAR), 323 MHz (GMRT; Swarup et al.\ 1991), and also at 0.5--7 keV (Chandra) have recently been published by Edler et al. (2022). Its integrated radio spectrum exhibits extreme curvature, with $\alpha_ {\rm 144-323 MHz} \approx - 4$ and $\alpha_{\rm 54-144 MHz} \approx - 2$. On the LOFAR 144 MHz image of this WAT (Fig.\ 2 of Elder et al. 2022, reproduced in Fig.\ 1 here), the brighter northern radio tail is traceable out to $\sim 700$ kpc from the parent galaxy. A ridge-line drawn through this radio tail clearly shows two bends (in opposite directions).  Note that these bends occur within the first half of the tail, at close to 200 kpc and 375 kpc distances from the nucleus of the WAT. Thereafter, the tail becomes straight. We have reproduced in Fig.\ 1 their 144 MHz LOFAR map with the ridge line of the northern tail
drawn, and the distances from the parent galaxy marked along the tail. 
In Fig.\ 2, adopted from the bottom panel of Fig.\ 4 of Edler et al.\ (2022), we show the spectral indices measured between 54 and 144 MHz (blue) and between 144 and 323 MHz (orange), as a function of distance from the parent galaxy, along the northern radio tail. 

The most striking feature evident from  Fig.\ 1  and Fig.\ 2 is that the two bends in the northern tail are also the only two locations at which the gradient of spectral index is seen to reverse its sign, with a localised spectral flattening ensuing from thereon, along the tail.
The trend is more conspicuous in the profile of $\alpha_{\rm 54-144 MHz}$, probably owing to its smaller error bars, and especially near the second bend (at $\sim 375$ kpc), which is the sharper of the two bends. 
A consistent trend is seen in the other spectral index profile, $\alpha_{\rm 144-323 MHz}$, along the northern tail, which is also shown in Fig.\ 2, and was derived in Elder et al., by combining a partly independent set of radio images (LOFAR HBA and GMRT). Note that, as expected, both spectral bends are also mirrored in the LOFAR radio fluxes measured along the northern tail at 144 MHz and 54 MHz (see the top panel in Fig.\ 4 of Elder et al.\ 2022). While a detailed investigation of the implications of the correlation noted here remains to be carried out, it does seem to add to the nascent body of evidence that bending of the collimated stream of synchrotron plasma can  play a substantive role in triggering {\it in-situ} particle acceleration, with detectable manifestations in some instances.   Conceivably, the bending results in an enhanced shear, triggering {\it in-situ} acceleration or re-acceleration of energetic particles (see Berezhko \& Krymskii 1981; Rieger \& Mannheim 2002, and references therein).  A collimated jet undergoing bending is subject to Kelvin-Helmholtz instabilities that also can induce the jet to flare (e.g., Loken et al.\ 1995; Lal 2021).

Several  other physical processes are capable of producing widespread reacceleration of electrons to energies that yield flatter synchrotron spectra at large distances from the AGN.  Essentially hydrodynamic instabilities of the Rayleigh-Taylor or Kelvin-Helmholtz types, if excited in the radio tails/lobes, can produce turbulent waves that  could drive second-order Fermi acceleration. In the case of the WAT in Abell 1033, de Gasperin et al. (2017) have focussed on the observed lack of spectral steepening along the relatively straight portion of the northern radio tail. This portion is the most distant (i.e., the oldest) one-third of the tail, as measured from the parent galaxy (Fig. 1). In the scenario preferred by de Gasperin et al, the near-spectral-constancy observed in this  straight portion of the tail is  probably caused by the distributed slow re-acceleration of particles in that region, due to turbulence generated by instabilities driven by interaction between the  straight portion of the tail and the (disturbed) ambient intra-cluster medium of that merging cluster. On the other hand, the onset of spectral flattening at both locations where the northern radio tail of this WAT undergoes bending, as pointed out here, strongly suggests that any {\it in situ} particle acceleration in the tail is probably  triggered at the points where the jet bends. 

In section 4.1 we have noted other known (albeit, somewhat less spectacular) examples of such a linkage. These are the WAT MRC $0600-399$ (Chibueze et al.\ 2021)as well as the XRGs PKS $2014-55$ (Cotton et al.\ 2020) and 3C 223.1 
(Gopal-Krishna \& Dabhade 2022). In addition, the apparently similar case of the radio galaxy 3C 40B actually pertains to radio filaments external to the radio lobes of a WAT. In a detailed study of this system, belonging to a poor cluster Abell 194, Rudnick et al.\ (2022) using their MeerKAT and LOFAR observations, have shown that the radio filaments extending eastward from the radio galaxy curve around the jet. They make a persuasive case that these magnetized filaments do not originate from the radio galaxy, but are instead generated via shear motions in the intracluster medium and are then stretched by interactions with the jet that amplify the magnetic fields within them.  This  process could reenergise the relativistic electrons in the filaments through betatron acceleration in the regions where the filaments  are stretched (Rudnick et al.\ 2022). Alternatively, as noted in section 4.1, when plasma flows are diverted via interaction with clumps of matter, or with significant external magnetic fields, the increased tension of the field lines can lead to magnetic reconnection between the jet and the compressed magnetic layer, as inferred from the MeerKAT observations of the WAT MRC $0600-399$ in the cluster Abell 3376 (Chibueze et al.\ 2021). Their magnetohydrodynamical simulations show that such a process can lead to efficient relativistic particle acceleration and their transport downstream (see also, Giri et al.\ 2022). A similar process may be operating in both known ``double boomerang'' type  XRGs, PKS 2014$-$55 and 3C 223.1, as proposed in Gopal-Krishna \& Dabhade (2022).

\subsection{Tail piece }

A key indication emerging from this work is that spectro-morphological details of the tails and lobes of radio galaxies  seem poised to unravel vital clues to the question of {\it in-situ} particle acceleration in the lobes of radio galaxies, a theme widely discussed and debated over the past half a century (sections 1 \& 2).  More specifically, the apparent spatial correlation between the lobe/tail bending and the onset of {\it in-situ} particle acceleration and consequent radio spectral flattening is providing important evidence.  This line of enquiry can now be vigorously pursued using the existing powerful high-resolution radio telescopes, JVLA, LOFAR, GMRT, and MeerKAT. 


\section*{Acknowledgements}

We thank Henrik Edler for permission to republish Fig.\ 1 and for providing the data used in plotting Fig.\ 2.  With the publication of this article, the authors will have co-authored 100 research publications. We take this opportunity to dedicate this article to the memory of Prof.\ Govind Swarup (1929 -- 2020) who seeded radio astronomy in India and steered its course for the first 5 decades. GK is thankful to the Indian National Science Academy (INSA) for an INSA Senior Scientist position which he currently holds.


\section*{References}
\bibliography{references}

\noindent Alexander P. 1987, MNRAS, 225, 27  

\noindent Alexander P., Leahy J. P. 1987, MNRAS, 225, 1 

\noindent Arshakian T. G.,   Longair M. S. 2000, MNRAS, 311, 846   

\noindent Athreya R. M., Kapahi V. K. 1998, JApA, 19, 63

\noindent Barai P., Wiita P. J. 2006, MNRAS, 372, 381 


\noindent Begelman M. C.,  Blandford R. D.,  Rees M. J. 1984, RvMP, 56, 255  

\noindent Berezhko E.~G., Krymskii  G.~F.  1981, Sov Astron Lett, 7, 352  

\noindent Blandford R. D.,  Rees M. J. 1974, MNRAS, 169, 395  

\noindent Blandford R., Meier, D.,  Readhead, A. 2019, ARA\&A, 57, 467 


\noindent Blundell K. M.,  Rawlings S. 1999, Nature, 399, 330  

\noindent Blundell K. M.,  Rawlings S. 2000, AJ, 199, 1111  

\noindent Blundell, K. M.,  Rawlings S. 2001, in  Laing R.~A., Blundell K.~M. eds, Particles and Fields in Radio Galaxies, ASP Conference Series., Volume 250, p.\  363 

\noindent Blundell K. M., Rawlings S.,   Willott C. J.  2002, NewAR, 46, 75  

\noindent Bolton J. G. 1956, PASP, 68, 477

\noindent Bovkun V.~P. 1976, Sov Astron,  19, 723

\noindent  Bridle A. H., Perley R. A. 1984, ARA\&A, 22, 319   

\noindent  Burbidge G. R. 1956, ApJ, 124, 4 

\noindent 

\noindent  Burch S. F. 1979, MNRAS, 186, 519


\noindent Capetti A.,  Zamfir S., Rossi P.  2002, A\&A, 394, 39

\noindent  Carilli C. L., Perley R. A., Dreher J. W., et al. 1991, ApJ, 383, 554

\noindent  Chibueze J. O., Sakemi H., Ohmura T. et al. 2021, Nature, 593, 47

\noindent  Christiansen W. 1969, MNRAS, 145, 327

\noindent Cohen M.~H. 1969, ARA\&A, 7, 619

\noindent Collins R.~A., Scott P.~F.  1969, MNRAS, 142, 371

\noindent Cotton W. D., Thorat K., Condon J. J. et al. 2000,   MNRAS, 495, 1271

\noindent de Gasperin F., Intema H. T., Shimwell T. W. et al. 2017, Sci Adv, 3, e10701634

\noindent Dennett-Thorpe J., Scheuer P. A. G., Laing R. A. et al. 2002, MNRAS, 330, 609

\noindent De Young D. S., Axford W. I., 1967, Nature, 216, 129

\noindent  Edler H. W., de Gasperin F., Brunetti G., et al. 2022, 666, 3


\noindent Eilek J. A.,  Arendt, P. N. 1996, ApJ, 457,  150 

\noindent Eilek J. A., Melrose D. B.,  Walker M. A. 1997, ApJ, 483, 282 

\noindent Ekejiuba I. E., Wiita P. J., Frazin R. A. 1994, ApJ, 434, 503

\noindent En{\ss}lin T. A., Gopal-Krishna  2001, A\&A, 366, 26

\noindent Fabian A. C. 2012, ARA\&A, 50, 455 

\noindent Fanaroff B. L.,  Riley J. M. 1974, MNRAS, 167, P31 


\noindent Getmantsev G. G.,  Ginzburg V. L. 1950, JETP, 20, 347 

\noindent Giri G., Vaidya B., \& Rossi P. 2022, A\&A, 662, 5 

\noindent Gopal-Krishna 1977, MNRAS, 181, 247

\noindent Gopal-Krishna 2021, Curr. Sci., 120, 1530

\noindent  Gopal-Krishna, Biermann P. L., Gergely L.,  Wiita P. J. 2012 , RAA, 12, 127 

\noindent Gopal Krishna,  Dabhade P. 2022, A\&A, 663, 8

\noindent Gopal-Krishna, Mhaskey M.,  Mangalam A. 2012, ApJ, 744, 31

\noindent Gopal-Krishna,  Swarup G. 1977, MNRAS, 178, 265

\noindent Gopal-Krishna,  Wiita P. J. 1987, MNRAS, 226, 531

\noindent Gopal-Krishna, Wiita P. J. 1990, A\&A, 236, 305

\noindent Gopal-Krishna, Wiita P. J.  1991, ApJ, 373, 325

\noindent Gopal-Krishna, Wiita P. J. 2005, in Saha S.~K., Rastogi V.~K. eds, 21st Century Astrophysics,  Anita Publications, New Delhi, p.\ 108 (arXiv:astro-ph/0409761) 

\noindent Gopal-Krishna, Wiita P. J.,  Saripalli L. 1989, MNRAS, 239, 173


\noindent Hardcastle M. J. 2013, MNRAS, 433, 3364

\noindent Hargrave P. J., Ryle M.  1974, MNRAS, 166, 305

\noindent Harris A. 1973 MNRAS, 163, 19P

\noindent Harwood J. J., Hardcastle M. J., Morganti R. et al. 2017, MNRAS, 469, 639


\noindent Hazard C., Mackey M. B.,  Shimmins A. J. 1963, Nature, 197, 1037


\noindent H{\"o}gbom J. A.  1979, A\&AS, 36, 173

\noindent 

\noindent Ishwara-Chandra C.\ H.,  Saikia D.\ J. 1999, MNRAS, 309, 100 

\noindent Jaffe W. J.,  Perola G. C.  1973. A\&A, 26, 423

\noindent Jamrozy M., Konar C., Machalski J.,  et al. 2008, MNRAS, 385, 1286

\noindent Jenkins C. L.,  Scheuer P. A. G. 1976, MNRAS, 174, 327

\noindent Jennison R. C.,  Das Gupta M. K. 1953, Nature, 172, 996


\noindent Jones T. W., Ryu D.,  Engel A. 1999,  ApJ, 512, 105


\noindent Kaiser C. R., Alexander P. 1997, MNRAS, 286, 215


\noindent Kardashev N. S.  1962, Sv Astr, 6, 317

\noindent Katz-Stone D. M.,  Rudnick L. 1997, ApJ, 488, 146

\noindent Kawakatu N., Kino M,  Nagai H.  2009, ApJ, 697, L173

\noindent Kellermann K. I.  1966, ApJ, 146, 621

\noindent Kellermann K. I., Pauliny-Toth I. I. K.,   Williams P. J. S. 1969, ApJ, 157, 1

\noindent Klamer I.~J., Ekers R.~D., Bryant J.~J., et al. 2006, MNRAS, 371, 852

\noindent Konar C.,  Hardcastle M. J. 2013, MNRAS, 436, 1595

\noindent  Lal D. V. 2021, ApJ, 915, 126

\noindent Lal D. V.,  Rao A. P. 2005, MNRAS, 356, 232

\noindent  Lal D. V.,  Rao A. P. 2007, MNRAS, 374, 1085

\noindent Lal D. V., Sebastian B., Cheung C. C., et al. 2019, AJ, 157, 195


\noindent Leahy J. P.,  Williams A. G. 1984, MNRAS, 210, 929

\noindent Liu R., Pooley G.,  Riley J. M. 1992, MNRAS, 257, 545

\noindent  Loken  C., Roettiger K., Burns J. O., Norman M.  1995, ApJ, 445, 80

\noindent Longair M. S., Ryle M.,  Scheuer P. A. G.  1973, MNRAS, 164, 243

\noindent MacDonald G. H., Kenderdine S.,  Neville A. C.  1968, MNRAS, 138, 259

\noindent Machalski J.,  Chyzy K. T.,  Stawarz A., et al. 2007, A\&A, 462, 43


\noindent Mahatma V. H., Hardcastle M. J.,  Croston J. H. 2020, MNRAS, 491, 5015

\noindent Mangalam A. V.,  Gopal-Krishna 1995, MNRAS, 275, 976



\noindent McKean J. P., Godfrey L. E. H., Vegetti S., et al. 2016, MNRAS, 463, 3143

\noindent Merritt D.,  Ekers R. D. 2002, Science, 297, 1310

\noindent Miley G. 1980, ARA\&A, 18, 165

\noindent Mills D. M.,  Sturrock P. A. 1970, ApL, 5, 105

\noindent Mitton S.,  Ryle M. 1969, MNRAS, 146, 221

\noindent Moffet A. T.  1966, ARA\&A, 4, 145

\noindent Murgia M., Fanti C., Fanti R., et al.  1999, A\&A, 345, 769

\noindent Myers S. T.,  Spangler S. R.  1985, ApJ, 291, 52

\noindent Nulsen P. E. J., Stewart G. C.,  Fabian A. C.  1984, MNRAS, 208, 185

\noindent O'Dea C. P., Daly R. A., Kharb P., et al. 2009, A\&A, 494, 471

\noindent Pacholczyk A. G  1970, Radio Astrophysics, W.H. Freeman, San Francisco 

\noindent Pacholczyk A. G.,  Scott J. S. 1976, ApJ, 203, 313 

\noindent Patra D., Joshi R.,   Gopal-Krishna  2023, MNRAS, 524, 3270

\noindent Perley R. A., Dreher J. W.,  Cowan J. J.  1984, ApJ, 285, L35

\noindent Pikelner S. B., 1953, Dokl Akad. Nauk. USSR, 88, 229

\noindent Rees M. J. 1971, Nature, 229, 312

\noindent Rieger  F.~M., Mannheim K. 2002, A\&A, 396, 833

\noindent Riley J. M,  Branson N. J. B. A.  1973,  MNRAS, 164, 271

\noindent Rottmann H.  2001,  PhD Thesis, U. Bonn 

\noindent Rudnick L., Br\"uggen M., Brunetti G., et al. 2022, ApJ, 935, 168

\noindent Rudnick L., Katz-Stone D. M.,  Anderson M. C.  1994, ApJS, 90, 955 

\noindent Ryle M. 1962, Nature, 194, 517

\noindent Ryle M. 1972, Nature, 239, 435

\noindent Safouris V., Subrahmanyan R., Bicknell G. V., et al.  2008, MNRAS, 385, 2117 

\noindent Scheuer P. A. G. 1962,  AuJPh, 15, 333

\noindent Scheuer P. A. G.  1974, MNRAS, 166, 513

\noindent Scheuer P. A. G.  1995,   MNRAS, 277, 331

\noindent Schmidt M. 1963, Nature, 197, 104  

\noindent Schmidt M. 1966, ApJ, 146, 7  


\noindent Shklovskii I. S., 1953, AZh, 30, 15

\noindent Shklovskii I. S., 1955, AZh, 32, 215 

\noindent Swarup G., Ananthakrishnan S.,  Kapahi V. K., et al. 1991, Cur. Sci., 60, 95

\noindent Swarup G., Sarma N. V. G., Joshi M. N., et al. 1971 Nature Phys Sci, 230, 185

\noindent  Swarup G.,  Thompson A. R.,  Bracewell R. N.  1963, ApJ, 138, 305

\noindent Taylor J. H.  1966,  Nature, 210, 1105


\noindent Taylor J. H., de Jong M. L. 1968, ApJ, 151, 33

\noindent Tribble P. C.  1993, MNRAS, 261, 57

\noindent Turland B. D. 1975, MNRAS, 172, 181

\noindent Turner R. J.,  Shabala S. S.  2015, ApJ, 806, 59

\noindent van Breugel, W. J. M. 1980, A\&A, 81, 265

\noindent van der Laan H.,  Perola G. C. 1969,   A\&A, 3, 468

\noindent van Weerden R. J., Andrade-Santos F., Dawson W. A., et al. 2017, Nature Astron, 1, id.\ 0005


\noindent von Hoerner S. 1964, ApJ, 140, 65


\noindent Wiita P. J.,  Gopal-Krishna  1990, ApJ, 353, 476  

\noindent Willis A. G.,  Strom R. G.  1978, A\&A, 62, 375

\noindent Winter A. J. B., Wilson D. M. A., Warner,P. J., et al.  1980, MNRAS, 192, 931

\noindent Zier, C., Biermann P. L. 2001, A\&A, 377, 23 

\end{document}